\documentclass[aps,pra,showpacs,reprint,superscriptaddress,floatfix]{revtex4-1}
\usepackage{graphicx}
\usepackage{longtable}
\usepackage{dcolumn}
\usepackage{multirow}
\usepackage{amssymb, amsmath}
\usepackage[colorlinks=true, citecolor=blue,linkcolor=blue]{hyperref}
\begin{document}
\title{ \emph{Ab} \emph{initio} calculations of spectroscopic constants and properties of BeLi$^+$}%
\author{Renu Bala}
\email[Electronic address: ]{rbaladph@iitr.ac.in}
\author{H. S. Nataraj}
\affiliation{ Department of Physics, Indian Institute of Technology Roorkee,
Roorkee - 247667, India}
\begin{abstract}
We have calculated the ground state spectroscopic constants and the molecular properties, of a molecular ion BeLi$^+$, such as dipole moment, quadrupole moment and dipole polarizability at different levels of correlation: many-body perturbation theory (MP2), coupled cluster method with single and double excitations (CCSD) and CCSD with perturbative triples (CCSD(T)). The correlation consistent polarized valence cc-pVXZ (X=D, T, Q) basis sets and also their augmented counterparts are used together with the non-relativistic and relativistic Hamiltonians. The results are extrapolated to the complete basis set limit (CBS) using exponential-Gaussian function. Thus, accurate and reliable results for BeLi$^+$ with the most conservative error estimates on them are reported.\\

Keywords: Spectroscopic constants, molecular properties, potential energy curve, dipole moment, quadrupole moment, dipole polarizability, BeLi$^+$.
\end{abstract}
\maketitle
%
\section{Introduction}
%
Understanding the complex structure of the molecules, neutral as well as ions, is an outstanding problem of molecular physics. 
With delicate investigations of such systems, one can obtain a wealth of information which can further be used to explore the various inter-related areas of research. To mention, cold and ultracold molecules can be used in the study of the controlled chemical reactions~\cite{Ospelkaus}, the variation of fundamental physical constants such as the proton-to-electron mass ratio, $\mu\equiv m_p/m_e$~\cite{Kajita,Kajita1,Flambaum}, the fine structure constant, $\alpha$~\cite{Flambaum}, and also in the search for the symmetry violating exotic property of an electron which hitherto has evaded detection, called the electric dipole moment (EDM)~\cite {Meyer,Hudson}, etc. A number of experimental techniques viz., buffer gas cooling~\cite{Weinstein} and electrostatic trapping~\cite{Bethem} for the formation of cold molecules, photoassociation (PA)~\cite{Sage} and Feshbach resonance~\cite{Ni} for ultracold molecules, and sympathetic cooling using laser cooled atomic ions,~\cite{Drewsen, Blythe, Staanum} for molecular ions, etc., have been developed to serve this purpose.\\ 
For over a decade, the neutral homonuclear and heteronuclear diatomic molecules of alkali- and alkaline-earth-metals have extensively been studied both theoretically~\cite {Fedorov, Deiglmayr, Geetha} and experimentally \cite {Haimberger,Ni,Kerman}. The calculation of spectroscopic constants, permanent dipole moments (PDM's), and vibrational states of electronic ground state of monohydride ions of group IIA, IIB and Yb have been reported in~\cite{Abe}. Recently also, ground and first excited state of the monohydrides of alkaline-earth-metals have been investigated theoretically using multi-reference configuration interaction (MRCI) method for the purpose of laser cooling ~\cite{Gao1}. The ground and low lying excited states of MgLi and MgLi$^+$ have been studied using full valence configuration interaction (FCI) and MRCI methods~\cite{Gao}. Quite recently, the structural, electronic and dipolar properties of CaLi$^+$ and SrLi$^+$ have also been reported in~\cite{Habli, Jellali}.\\
Although for a simple system like BeLi$^+$ there is no available experimental data, to the best of our knowledge. But there exist several  calculations in the literature, such as, the calculation of spectroscopic constants of 1-2\,$^1\Sigma^+$ and 1\,$^3\Sigma^+$ using self-consistent field (SCF) method by Safonov \emph{et al.}~\cite{Safonov}, stability and physicochemical reaction of this molecular species has been studied by Nicolaides \emph{et al.}~\cite{Nicolaides}, the calculation of ground state spectroscopic constants at MP2 (full) level by Boldyrev \emph{et al.}~\cite {Boldyrev} using the Gaussian 92 program, calculations of the ground state and low-lying $^1\Sigma^+$ electronic states of BeLi$^+$ by Farjallah \emph{et al.}~\cite {Farjallah} using FCI, Rakshit~\emph{et al.}~\cite{Rakshit} reported the ground and first excited state potential energy curves (PEC's) to study the atom-ion collision at low energy and to predict the feasibility of the formation of cold molecular ion by photoassociation, ground and six low-lying excited states have been studied by Sun~\emph{et al.}~\cite{Sun} using MRCI + Q and multi-reference averaged quadratic coupled-cluster (MRAQCC) methods including Davidson correction. Recently, \emph{ab initio} study of ground and low-lying excited states have been done by You \emph{et al.}~\cite{You} using MRCI method. Very recently, Ghanmi~\emph{et al.}~\cite{Ghanmi} performed the theoretical study of 44 excited electronic states of BeLi$^+$ molecular ion using pseudo-potential approach. In this paper, they have reported the PEC's, spectroscopic constants, dipole moment curves, and transition dipole moment curves. In the present work, for achieving reliable accuracy we have employed the higher levels of correlation methods and also we have used a successive hierarchy of three optimized basis set, so that the results can be extrapolated to the CBS limit to bring the saturation due to the size of the basis. Thus, we have calculated the spectroscopic constants, potential energy curves and dipole moment curves of BeLi$^+$ systematically and more accurately. In addition, we have calculated and reported the dipole polarizability and the quadrupole moment of BeLi$^+$ ion for the first time, known to our knowledge.\\  
The present paper is divided into four sections. After the introduction, Section~\ref{section-2} briefly reports the methodology involved in the calculations, Section~\ref{section-3} details the discussions on the calculated results and Section~\ref{section-4} summarizes the current work.
%
\section{Methodology}
\label{section-2}
%
CFOUR~\cite{CFOUR} and DIRAC15~\cite{DIRAC} software suites are used to carry out the non-relativistic and relativistic calculations, respectively, of the spectroscopic constants: equilibrium bond lengths ($R_e$), dissociation energies
\begin{table*}[ht]
\begin{ruledtabular}
\begin{center}
\caption{\label{table-I} The computed spectroscopic constants for the ground state of BeLi$^+$, in the non-relativistic case and the available results in the literature.}
\begin{tabular}{cccccccc}
Basis & Method & $R_e$~({\emph {au}}) & $D_e$ (cm$^{-1}$) & $\omega$$_e$(cm$^{-1}$) & $\omega$$_ex_e$(cm$^{-1}$) & $B_e$ (cm$^{-1}$) & $\alpha_e$ (cm$^{-1}$)\\
  \hline 
  DZ & SCF & 5.006 & 5068& 305 & 5.01& 0.608 & 0.0138 \\
     & MP2 & 4.966 & 4994 & 320 & 4.80 & 0.618 & 0.0130 \\
     & CCSD & 4.957 & 4644 & 318 & 4.84 & 0.621 & 0.0131 \\
     & CCSD(T)& 4.955 & 4655 & 318 & 4.93 & 0.621 & 0.0131 \\
     \hline
  TZ & SCF & 5.008 & 5105 & 302 &5.16 & 0.608 & 0.0139\\
     & MP2 & 4.947 & 5145& 324 & 4.82 & 0.623 & 0.0129 \\
     & CCSD  & 4.942 & 4799 & 319 & 4.84 & 0.624 & 0.0131 \\
     & CCSD(T)& 4.940 & 4812 & 319 & 4.82 & 0.625 & 0.0131 \\
     \hline
  QZ & SCF & 5.004 & 5116 &300 & 5.15 & 0.609 & 0.0144\\
     & MP2 & 4.930 &5222 & 326& 4.95 & 0.627 & 0.0132 \\
     & CCSD & 4.927 & 4859 & 321 & 4.86 & 0.628 & 0.0134 \\
     & CCSD(T)& 4.923 & 4868 &322 & 4.85 & 0.629 & 0.0133 \\
     \hline
\textbf{CBS}  & SCF &5.001 & 5122 & 299& 5.14& 0.610 & 0.0147 \\
     & MP2 & 4.919 & 5269 & 327 & 5.04 & 0.630 & 0.0134 \\
 & CCSD & 4.917 & 4894 & 322 & 4.87 & 0.631 & 0.0136\\
 & \textbf{CCSD(T)} & \textbf{4.912} & \textbf{4900} & \textbf{324} & \textbf{4.87} & \textbf{0.631} & \textbf{0.0134} \\
   \textbf{Error bar}  & &\textbf{$\pm$0.005} & \textbf{$\pm$6} & \textbf{$\pm$2} & \textbf{$\pm$0.0006} & \textbf{$\pm$0.0009} & \textbf{$\pm$0.0002}\\
   \hline
 \multicolumn{2}{c}{\textbf{published works}} &\\
 & SCF~\cite{Safonov} & 5.027 & 5001 & 320 & 8.2 & 0.601 & $-$ \\
 & \cite{Nicolaides} & 4.920  &  $-$ & $-$ & $-$ & $-$   & $-$ \\
 & MP2~\cite{Boldyrev}& 4.968 & $-$ & 320 & $-$ & $-$& $-$\\
 & MP4$^{*}$~\cite{Boldyrev}& $-$ & 4757 & $-$ & $-$ & $-$ & $-$\\
 & QCISD(T)$^{**}$~\cite{Boldyrev}& $-$ & 4617 & $-$ & $-$ & $-$ & $-$\\
 & FCI~\cite{Farjallah}& 4.96 & 4849 & 321.2 & 6.48 & 0.6395& $-$ \\
 & CCSD(T)~\cite{Sun} & 4.964 & $-$ & 315.5 & $-$ & 0.623 & $-$\\
 & MRCI~\cite{You} & 4.913 & 4903.6 & 318.4 & 4.310 & 0.6173 & $-$\\
 & FCI~\cite{Ghanmi} & 4.940 & 4862 & 323.7 & 5.45 & 0.6395 & $-$\\
 & MRD-CI$^{***}$~\cite{Pewestorf} & 5.083   &  $-$ & $-$ & $-$ & $-$   & $-$ \\
\end{tabular}
\begin{flushleft}
* MP4 - Fourth-order many-body perturbation theory, ** QCISD(T) - Quadratic configuration interaction including singles and doubles with partial triples, *** MRD-CI - Multireference Determinant configuration interaction.
\end{flushleft}
\end{center}
\end{ruledtabular}

\end{table*}
($D_e$), harmonic frequencies ($\omega_e$), anharmonic constants ($\omega_ex_e$), equilibrium rotational constants ($B_e$),  and the molecular properties: dipole moments ($\mu_e$), quadrupole moments ($\Theta_{zz}$) and components of static dipole polarizability ($\alpha_{\parallel}\, =\,\alpha_{zz}$ and $\alpha_{\bot}\,=\, \alpha_{xx}$ or $\alpha_{yy}$, should $z$-axis be the internuclear axis of the molecule) at MP2, CCSD and CCSD(T) level of correlation. The uncontracted cc-pVXZ and aug-cc-pVXZ basis sets of Dunning \emph {et al.}~\cite {Dunning} with X=[D, T, Q], available in the DIRAC15 package, are used in the present calculations. The nuclear mass of 9.0121822 {\emph {au}} for Be and of 7.0160030 {\emph {au}} for Li have been used. Further, the point nuclear distribution and C$_{2v}$ molecular point group symmetry are considered in our calculations. All the electrons are kept active both for diatomic constants and also for property calculations. Furthermore, the energies and PDM's are calculated for the range 2-30 {\AA} with a step size of 1 {\AA} and around the equilibrium point a finer step size of 0.001 {\AA} is considered. The maximum distance of 30 {\AA} is chosen based on the saturation of energies limited to the threshold set. The dissociation energies are evaluated by taking the difference between the energies at equilibrium bond length and those at a distance of 30 {\AA}.\\
For non-relativistic calculations, the basis sets have been taken from the EMSL library~\cite{EMSL}. The harmonic frequencies and anharmonic constants are calculated using the second-order vibrational perturbation theory, i.e., with VPT2 keyword in the CFOUR package. All calculations of the diatomic constants and of properties are carried out at zero frozen-core orbitals level and also without cutting-off any virtual orbitals. The average polarizabilities ($\bar{\alpha}$) and anisotropic polarizabilities ($\gamma$ ) are obtained, respectively, using: 
  \begin{eqnarray}{}
\bar{\alpha}\,=\,(\alpha{_\parallel }\,+\,2\,\alpha{_\bot})/3, 
\end{eqnarray}
and
 \begin{eqnarray}{}
\gamma\,=\,\alpha{_\parallel}\,-\,\alpha{_\bot}.
\end{eqnarray} 
The dipole moment of the system is calculated via first-order derivative of energy with respect to the electric field as,
 \begin{eqnarray}{}
 \mu\,=\,-\left(\frac{dE}{d\epsilon}\right)_{\epsilon=0}
 \end{eqnarray}
The $z$-component of the quadrupole moment, $\Theta_{zz}$ is related to the other diagonal components by the relation,
\begin{eqnarray}{}
   \Theta_{zz}\,=\,-(\Theta_{xx}\,+\,\Theta_{yy})
\end{eqnarray}
and for linear molecules, $\Theta_{xx}\,=\,\Theta_{yy}$. Therefore,
\begin{eqnarray}{}
 \Theta_{zz}\,=\,-2\Theta_{xx}
\end{eqnarray}
For the relativistic case, the Dirac-Fock-Coulomb Hamiltonian is used with the DIRAC15 package. The contribution from the (SS$|$SS) integrals is taken in an approximate manner, as suggested in ~\cite{Visscher}, by including an interatomic (SS$|$SS) correction. The energy calculations at MP2, CCSD and CCSD(T) levels are carried out with the RELCCSD module, keeping a cut-off energy of $10E_h$ for limiting the higher virtual orbitals. The diatomic constants are calculated by the second-order differentiation of the potential energy curves. The harmonic frequencies, anharmonic constants and equilibrium rotational constants, respectively, are obtained using: 
%
%
\begin{table*}[ht]
\begin{ruledtabular}
\begin{center}
\caption{\label{table-II} The computed spectroscopic constants for the ground state of BeLi$^+$, in the relativistic case..}

\begin{tabular}{cccccccc}
Basis & Method &$R_e$~({\emph {au})} & $D_e$ (cm$^{-1}$) & $\omega$$_e$(cm$^{-1}$) & $\omega$$_ex_e$(cm$^{-1}$) & $B_e$ (cm$^{-1}$) \\
  \hline 
  DZ & SCF & 5.006 & 5067 & 303 & 4.53 & 0.609 \\
     & MP2 & 4.964 & 5009 & 327 & 5.35 & 0.619 \\
     & CCSD & 4.955 & 4660 & 310 & 5.17 & 0.621 \\
     & CCSD(T)& 4.953 & 4671 & 323 & 5.60 & 0.622 \\
     \hline
  TZ & SCF & 5.006 & 5104 & 306 & 4.59 &  0.609 \\
     & MP2 & 4.947 & 5196 & 323 & 5.03 & 0.624 \\
     & CCSD & 4.944 & 4847 & 319 & 5.26 & 0.624 \\
     & CCSD(T)& 4.940 & 4860 & 311 & 5.00 & 0.625 \\
     \hline
  QZ & SCF & 5.004 & 5115 & 356 & 6.06 & 0.610 \\
     & MP2 & 4.919 & 5305 & 334 & 5.25 & 0.631 \\  
     & CCSD & 4.913 & 4955 & 318 & 5.10 & 0.632 \\
     & CCSD(T)& 4.910 & 4965 & 331 & 5.53 & 0.633 \\
     \hline
\textbf{CBS}   & SCF & 5.002 & 5121 & 390 & 7.05 & 0.611 \\
   & MP2 & 4.901 & 5372 & 342 & 5.41 & 0.635\\
    & CCSD & 4.892 & 5022 & 317 & 4.99  & 0.637\\
                & \textbf{CCSD(T)} & \textbf{4.890} & \textbf{5019} & \textbf{345} & \textbf{5.91} & \textbf{0.638}\\
\textbf{Error bar}& & \textbf{$\pm$0.002} &  \textbf{$\pm$3} &  \textbf{$\pm$28} &  \textbf{$\pm$0.92} &  \textbf{$\pm$0.001}\\
\end{tabular}
\begin{flushleft}
\end{flushleft}
\end{center}
\end{ruledtabular}
\end{table*}

\begin{eqnarray}{}
\omega_e\,=\,\frac{1}{2\pi}\,\sqrt{\frac{K}{\mu}},
\end{eqnarray}
\begin{eqnarray}{}
\omega_ex_e\,\cong\,\frac{\hbar\omega_e^2}{4D_e},
\end{eqnarray} 
and
\begin{eqnarray}{}
 B_e\,=\,\frac{h}{8\pi^2c\mu R_e^2},
\end{eqnarray}
where $K$, $\mu$ and $c$ represent the force constant, the reduced mass of the constituent atoms of a molecule and the speed of light, respectively. 
The calculated results are extrapolated to CBS limit using a function of the form \cite{Peterson, Feller},
\begin{eqnarray}{}
\label{CBS}
f(x)=f_{CBS}+B\, e^{-(x-1)}+C\, e^{-(x-1)^2}
\end{eqnarray} 
where $B$ and $C$ are constant parameters, $x = 2, 3, 4$ is the cardinal number for basis sets DZ, TZ and QZ, respectively, $f(x)$ is the property calculated with the basis set characterized by cardinal number $x$ and $f_{CBS}$ is the complete basis set limit for the property of interest.\\ 
The estimation of the uncertainty for the calculations, in general, is a little tricky; there can be two possible error sources: one arising due to the choice or the size of the basis set and the other arising from the negligence of higher order correlation effects beyond what is considered in the present work. As we have done a series of calculations with a successive hierarchy of optimized basis sets, and extrapolated the results systematically, to the CBS limit, we least expect the error due to the choice of the basis set. As the contributions from the higher order correlation effects are not expected to be larger than that of the leading order triples considered perturbatively in our work, we have quoted the entire contribution from the latter, viz. CCSD(T) - CCSD at CBS level, as the maximum possible error bar ($\Delta$) on our results. 
Thus, we recommend the final result, $f_{final}$, of the calculated property as,  
\begin{eqnarray}{}
f_{final}\,=\,f_{CCSD(T)}\,\pm\,(\arrowvert\Delta\arrowvert)
\end{eqnarray}
where $f_{CCSD(T)}$ is the CBS value of the property of interest at CCSD(T) level.
%
\begin{figure}[h]
\includegraphics[width=\columnwidth]{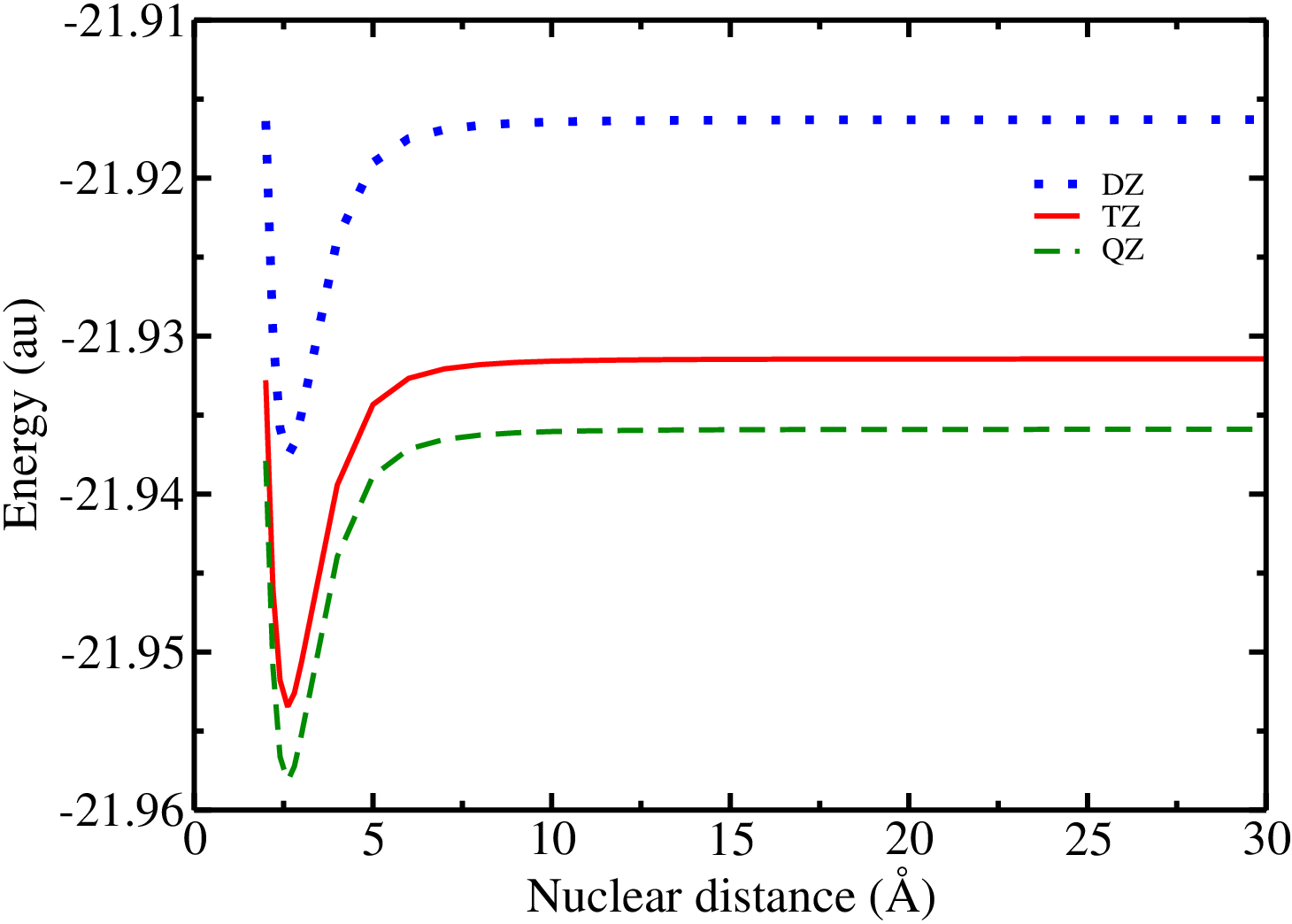}
\caption{\label{FIG.1.}(colour online) Potential energy curves for the ground state of BeLi$^+$, at CCSD(T) level of correlation, in the non-relativistic case.}
\end{figure}
%
\section{Results and discussion}
\label{section-3}
\begin{figure}[ht]
\includegraphics[width=\columnwidth]{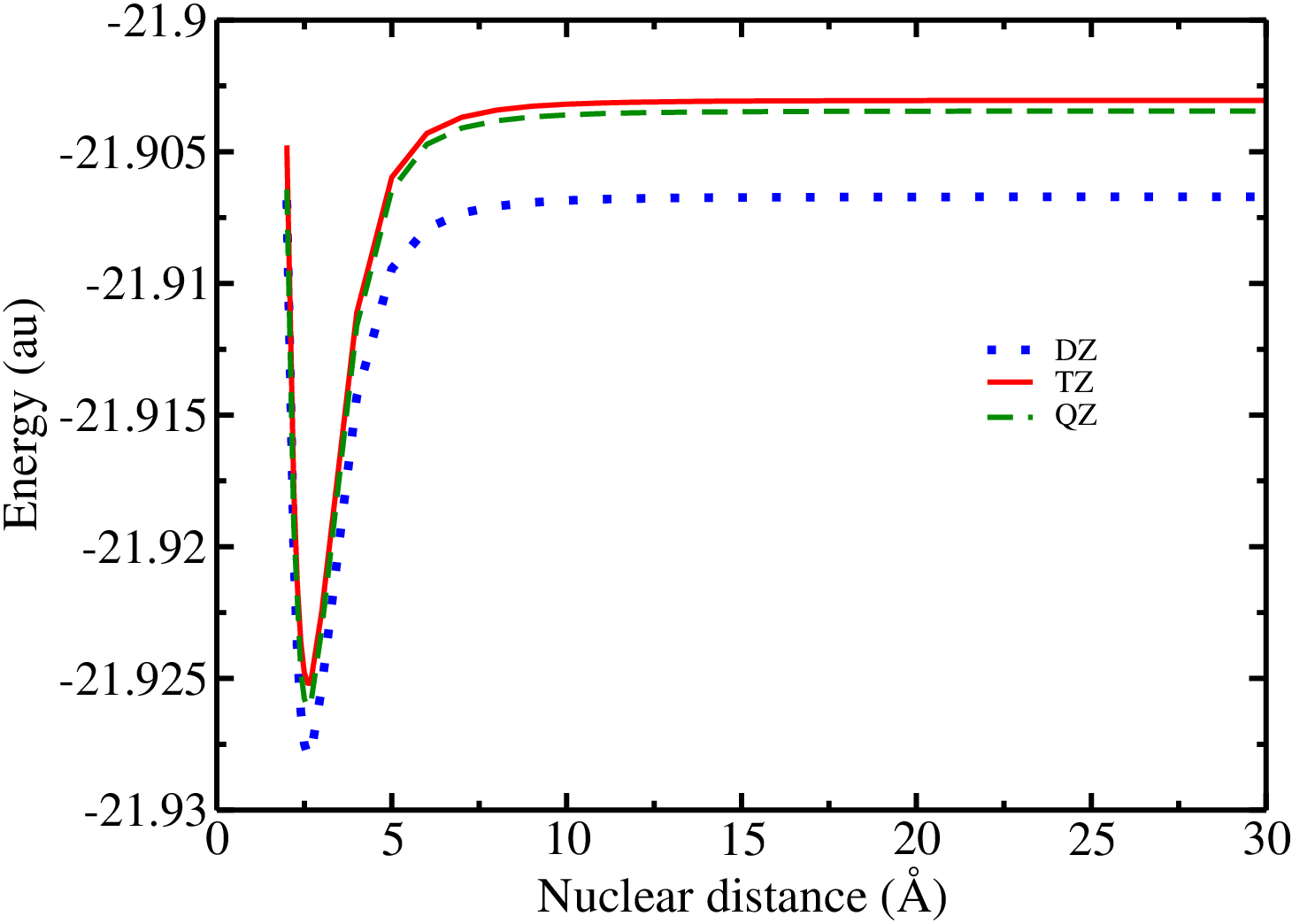}
\caption{\label{FIG.2.}(colour online) Potential energy curves for the ground state of BeLi$^+$, at the CCSD(T) level of correlation, in the relativistic case.}
\end{figure}

\begin{figure}[h]
\includegraphics[width=\columnwidth]{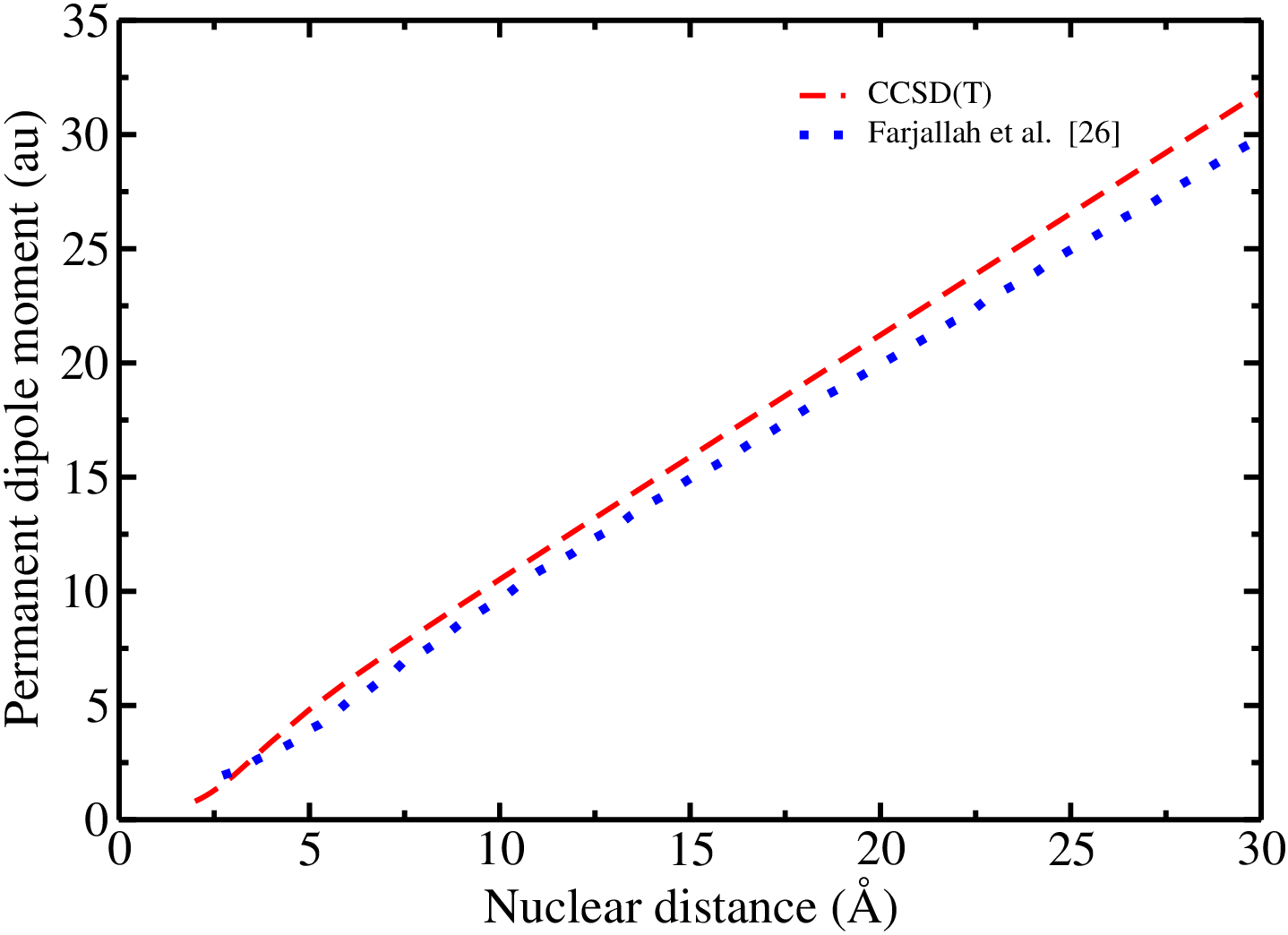}
\caption{\label{FIG.3.}(colour online) Permanent dipole moment curve using QZ basis set at CCSD(T) level of correlation.}
\end{figure}

%
Figures~\ref{FIG.1.}-\ref{FIG.2.} show the potential energy curves for the ground state of BeLi$^+$, obtained at the CCSD(T) level of correlation, using the non-relativistic and the relativistic Hamiltonian, respectively. Although we do not observe a clear trend between the two, the non-relativistic PEC's seem to be lower in energy.\\
The spectroscopic constants computed, using the non-relativistic Hamiltonian, at various levels of correlation considered in this work are shown in Table~\ref{table-I}. In addition, the results extrapolated to CBS limit are also shown together with the results available from the literature. Our results for the spectroscopic constants shows a good agreement with the available theoretical results~\cite{Safonov, Nicolaides, Boldyrev, Farjallah, Sun, You, Ghanmi, Pewestorf}. The difference between our CBS values at SCF level and those quoted in \cite{Safonov} is not more than 7\% in all the spectroscopic constants but anharmonic constant. The latter, which depends on other diatomic constants, shows a larger deviation. The MP2 level results of Ref.~\cite{Boldyrev} also compare quite well with our results at the similar level of approximation. The results reported in ~\cite{Farjallah}, calculated at FCI level and our recommended CCSD(T) results, highlighted in bold fonts, agree extremely well. According to the spectroscopy study of Sun \emph{et al.}~\cite{Sun} at CCSD(T) level, BeLi$^+$ molecular ion has equilibrium distance $R_e\,=\,4.964~au$, harmonic frequency $\omega_e\,=\,315.5~cm^{-1}$, and rotational constant $B_e\,=\,0.623~cm^{-1}$, which are close to our result of 4.912 $au$, 324 $cm^{-1}$ and 0.631 $cm^{-1}$ respectively, at same level of correlation. Except the anharmonic constant, which is 0.56 $cm^{-1}$ larger than that of You \emph{et al.}~\cite{You}, and 0.58 $cm^{-1}$ smaller than that of Ghanmi \emph{et al.}~\cite{Ghanmi}, the difference between our CBS results at CCSD(T) level and those reported in Ref.~\cite{You, Ghanmi} at MRCI and FCI level of correlation respectively, is less than 3\% in all diatomic constants.\\
The results of the similar calculations, performed for the sake of completion, using the relativistic Hamiltonian are given in Table~\ref{table-II}. As is obvious, due to the small charge of Be and Li, we did not observe an appreciable difference between the relativistic and the non-relativistic results in all the spectroscopic constants. \\
\begin{table*}[ht]
\begin{ruledtabular}
\begin{center}
\caption{\label{table-III} Dipole moment, quadrupole moment, components of dipole polarizability at equilibrium point, and polarizability at super-molecular limit, for the ground state of BeLi$^+$.}
\begin{tabular}{c c c c c c c c c}
Basis & Method &  $\arrowvert\mu_e\arrowvert$~({\emph {au}}) & $\Theta_{zz}$~({\emph {au}}) & $\alpha$$_{\parallel }$~({\emph {au}}) & $\alpha$$_{\bot }$~({\emph {au}}) & $\bar{\alpha}$~({\emph {au}}) & $\gamma$~({\emph {au}}) & $\alpha$$_{100}$ ({\emph {au}}) \\
  \hline
  DZ & SCF &  1.274 &  12.642 & 46.086  & 36.843 & 43.005 & 9.243 & 44.586 \\
     & MP2 &  1.335 & 12.713 & 43.881 & 34.285 & 37.483 & 9.596 & 40.403\\
     & CCSD &  1.449 & 11.732 & 42.010 & 32.071 & 35.384 & 9.939 & 35.921\\
     & CCSD(T)&   1.450 & 11.712 & 41.942 & 32.006 & 35.318 & 9.936 & 35.838\\
     \hline
  TZ & SCF & 1.276 & 12.622 & 45.941 & 36.713 & 42.865 & 9.228 & 45.520\\
     & MP2 &  1.310 & 12.130 & 43.652 & 34.306 & 37.421 & 9.346 & 41.348\\
     & CCSD &   1.421 & 11.711 & 42.223 & 32.347 & 35.639 & 9.876 & 37.012\\
     & CCSD(T)&   1.422 & 11.687 & 42.143 & 32.273 & 35.563 & 9.870 & 36.911\\
     \hline
  QZ & SCF &  1.274 & 12.621 & 45.895 & 36.745 & 42.845 & 9.150 & 45.721\\
     & MP2 &  1.291 & 12.097 & 43.499 & 34.407 & 37.438 & 9.092 & 41.572\\ 
     & CCSD &   1.405 & 11.680 & 42.197 & 32.466 & 35.710 & 9.731 & 37.191\\
     & CCSD(T)&   1.406 & 11.645 & 42.087 & 32.363 & 35.604 & 9.724 &  37.037 \\
     \hline
\textbf{CBS}   &SCF &  1.273 & 12.621 & $-$ & $-$ & 42.836 & 9.098 & 45.826 \\
   & MP2 & 1.279 & 12.095 & $-$ & $-$ & 37.452 & 8.929 & 41.691 \\
   & CCSD &  1.395 & 11.660 & $-$ & $-$ & 35.749 & 9.635 & 37.274\\
   & \textbf{CCSD(T)} &  \textbf{1.396} & \textbf{11.617} & $-$ & $-$ & \textbf{35.623} & \textbf{9.628} & \textbf{37.085} \\
\textbf{Error bar}  & & \textbf{$\pm$0.001} & \textbf{$\pm$0.043} & $-$ & $-$ & \textbf{$\pm$0.126} & \textbf{$\pm$0.007} & \textbf{$\pm$0.189} \\
\end{tabular}
\begin{flushleft}
\end{flushleft}
\end{center}
\end{ruledtabular}
\end{table*}
The computed results, using the non-relativistic Hamiltonian, of molecular properties such as dipole moments, quadrupole moments, components of static dipole polarizability, isotropic and anisotropic polarizabilities, and parallel-component of the dipole polarizabilities at super molecular limit (100~\emph{au}) symbolized as $\alpha_{100}$, are tabulated in Table~\ref{table-III}. Knowledge of dipole moment is useful for the study of electric dipole-dipole interactions, quantum computing, etc. \cite{DeMille, Menotti}. Figure~\ref{FIG.3.} shows the PDM as a function of internuclear distance $R$ at CCSD(T) level of correlation using cc-pVQZ basis set. The dipole moment curve shows almost linear behaviour from 6 {\AA} to 30 {\AA}. In this figure, we have compared our results with the results reported by Farjallah \emph{et al.}~\cite{Farjallah}. The excess of negative charge on the lighter atom is confirmed by the negative values of dipole moment, with the assumption that the orientation of the interatomic axis is from lighter to heavier element. We have, however, plotted the absolute values of dipole moment against $R$. The dipole polarizability describes the behaviour of a molecule in the presence of electric field and hence, the dipole polarizability is an important property for the study of the optical properties of materials, intermolecular forces, etc. As we move towards the higher basis set, most of the results for spectroscopic constants and molecular properties converge.\\
The augmentation of the basis sets with diffuse functions does not appreciably alter the results of the spectroscopic constants and the molecular properties. The differences in our final results between the augmented and un-augmented basis sets is about, 0.001 {\emph {au}} in $R_e$, 1 cm$^{-1}$ in $D_e$ and $\omega_e$, 0.2 cm$^{-1}$ in $\omega_ex_e$, 0.0003 cm$^{-1}$ in B$_e$ 0.0001 cm$^{-1}$ in $\alpha_e$,  0.0004 {\emph {au}} in $\mu_e$ and $\Theta_{zz}$, 0.066 {\emph {au}} in $\bar\alpha$, and 0.051 {\emph {au}} in $\gamma$, at the CCSD(T) level. However, the polarizability results, with and without augmentation, in the super-molecular limit seem to have a noticeable difference. At the CCSD(T) level, the CBS value of $\alpha_{100}$ = (37.944 $\pm$ 0.167)\emph{au} with augmented basis set, while $\alpha_{100}$ = (37.085 $\pm$ 0.189)\emph{au} without augmentation. The former is significantly close to the the sum of polarizabilities of the Be \cite{Peter} and Li$^+$ \cite{Miadokova}, $\alpha_A \simeq \alpha_{Be}\,+\,\alpha_{Li^+}\,=\,37.71\,+\,0.191\, =\,37.901{\emph {au}}$. Due to the the unavailability of experimental value of polarizability for Li$^+$ in the literature, we have taken its calculated value at the CCSD(T) level from \cite{Miadokova} and combined with the experimental value for Be.\\

%
%
%
\section{\label{Summary}Summary}
\label{section-4}
%
In summary, we have carried out the \emph{ab initio} calculations of the spectroscopic constants and the molecular properties of the ground state ($^1\,\Sigma^+ $) of BeLi$^+$ molecular ion dissociating into Be + Li$^+$. The objective of this paper is to systematically improve the accuracies of the hitherto existing results. The saturation of the results with respect to the size of the basis sets and the higher levels of correlations are investigated. Our findings of diatomic constants and dipole moment are in reasonably good agreement with the available results in the literature. Other properties such as the dipole polarizability and the quadrupole moment for this ion are reported for the first time, in this work, to the best of our knowledge. We have reported very conservative error estimates on our results by considering the entire perturbative triples contribution as the maximum possible error.
\begin{center}
{\bf {ACKNOWLEDGMENTS}}
\end{center}
We are indebted to the authors of CFOUR and DIRAC15 packages for licencing us to use their resource-rich codes and also for the help provided through user groups and forums. All calculations reported in this work are performed on the computing facility in the Department of Physics, IIT Roorkee.
\end{document}